\documentclass{article} % For LaTeX2e
\usepackage{iclr2027_conference,times}

\usepackage{amsmath,amsfonts,bm}

\def\eqref#1{equation~\ref{#1}}
\def\1{\bm{1}}

\DeclareMathAlphabet{\mathsfit}{\encodingdefault}{\sfdefault}{m}{sl}
\SetMathAlphabet{\mathsfit}{bold}{\encodingdefault}{\sfdefault}{bx}{n}

\usepackage{longtable}
\usepackage{multirow}

\usepackage{amsmath}
\usepackage{amssymb}
\usepackage{mathtools}
\usepackage{amsthm}
\usepackage{xcolor}
\usepackage[most]{tcolorbox}
\newtcolorbox{evidencebox}[1]{
  enhanced, colback=gray!5, colframe=black, sharp corners,
  boxrule=0.2mm, boxsep=1mm, left=1mm, right=1mm,
  top=1mm, bottom=1mm, before skip=8pt, after skip=8pt,
  before upper={\textbf{#1}\par\smallskip}
}
\theoremstyle{plain}
\newtheorem{definition}{Definition}
\usepackage{booktabs}
\usepackage{wrapfig}
\usepackage{environ}
\newsavebox{\measuredwrapbox}
\newcommand{\measuredwrapclearance}{0.5\baselineskip}
\newsavebox{\measuredheadingbox}
\newcommand{\wrapheading}[1]{%
  \par\global\setbox\measuredheadingbox=\vbox{\hsize=\textwidth #1\par}%
}
\makeatletter
\newcommand{\placemeasuredwrap}{%
  \Needspace{\dimexpr\ht\measuredwrapbox+\dp\measuredwrapbox+\ht\measuredheadingbox+\dp\measuredheadingbox+2\intextsep+\measuredwrapclearance\relax}%
  \ifvoid\measuredheadingbox\else\noindent\box\measuredheadingbox\par\fi
  \begin{wrapfloat}{\measuredwraptype}{\measuredwrapside}{\measuredwrapwidth}%
  \vspace{-0.5\intextsep}%
  \usebox{\measuredwrapbox}\end{wrapfloat}%
}
\NewEnviron{autowraptable}[2]{%
  \par\global\setbox\measuredwrapbox=\hbox{\begin{minipage}{#2}\def\@captype{table}\setlength{\abovecaptionskip}{3pt}\setlength{\belowcaptionskip}{3pt}\BODY\end{minipage}}%
  \gdef\measuredwrapclearance{0pt}\gdef\measuredwraptype{table}\gdef\measuredwrapside{#1}\gdef\measuredwrapwidth{#2}%
  \aftergroup\placemeasuredwrap
}
\NewEnviron{autowrapfigure}[2]{%
  \par\global\setbox\measuredwrapbox=\hbox{\begin{minipage}{#2}\def\@captype{figure}\BODY\end{minipage}}%
  \gdef\measuredwrapclearance{0.5\baselineskip}\gdef\measuredwraptype{figure}\gdef\measuredwrapside{#1}\gdef\measuredwrapwidth{#2}%
  \aftergroup\placemeasuredwrap
}
\makeatother
\usepackage{capt-of}
\usepackage{array}
\usepackage{needspace}

\usepackage{algorithm}
\usepackage{algpseudocode}

\usepackage[hidelinks]{hyperref}
\usepackage{tikz}
\usepackage{url}
\usetikzlibrary{arrows.meta,positioning}

\newcommand{\papertablestyle}{%
    \small
    \renewcommand{\arraystretch}{1.12}%
    \setlength{\tabcolsep}{5pt}%
    \setlength{\heavyrulewidth}{0.7pt}%
    \setlength{\lightrulewidth}{0.35pt}%
}

\newcommand{\method}{$\mathtt{FragToken}\space$}
\newcommand{\methodbold}{$\bm{\mathtt{FragToken}}$}

\title{\method{}: Amplifying LLM Inference Costs through Noncanonical Token Generation
}

\iclrfinalcopy 
\author{
\textbf{Zihan Wang}$^{1}$ \qquad
\textbf{Rui Zhang}$^{1}$ \qquad
\textbf{Xinyuan Qian}$^{1}$ \\[2pt]
\textbf{Qingchuan Zhao}$^{2}$ \qquad
\textbf{Hongwei Li}$^{1}$ \qquad
\textbf{Guowen Xu}$^{1}$ \\[6pt]
{\normalfont\small $^{1}$University of Electronic Science and Technology of China} \\
{\normalfont\small $^{2}$City University of Hong Kong}
}

\begin{document}
\raggedbottom % Do not stretch paragraph and float gaps to fill the page.

\maketitle

\begin{abstract}
% The continued scaling of large language models (LLMs) improves model capabilities but also substantially increases inference latency and resource consumption, making resource-consumption attacks increasingly consequential. 
As large language model (LLM) inference becomes increasingly expensive, resource-consumption attacks pose a growing threat to model providers.
Existing attacks typically amplify cost by inducing abnormally long or repetitive outputs on attacker-controlled or triggered requests, making them easier to detect and limiting their deployment-wide impact when benign traffic dominates.
% In this work, we uncover a previously overlooked token-level attack surface: the same textual content can be generated through multiple token sequences with substantially different lengths.
In this work, we uncover a previously overlooked token-level attack surface arising from the many-to-one mapping from token sequences to decoded text.
% Although modern LLMs strongly favor the canonical token sequences defined by their tokenizers, steering generation toward these alternatives increases autoregressive decoding cost without proportional growth in visible output.
Although standard LLMs predominantly generate the canonical token sequences induced by their tokenizers, the same text can also be represented by substantially longer non-canonical sequences. 
This representational flexibility exposes a new avenue for resource-consumption attacks: an attacker can train the model to favor such sequences, systematically increasing the number of autoregressive decoding steps without a proportional increase in visible response length.
However, we empirically find that directly maximizing token fragmentation substantially degrades model utility, producing conspicuous answer-quality failures that undermine attack stealthiness.
To address this challenge, we propose \method{}, a training-time framework that combines source-model self-distillation, capacity-aware filtering and budgeting, and BPE-Aligned Merging to induce fragmented generation under ordinary prompts while largely preserving model utility.
We evaluate \method{} on four LLMs across three benchmarks. 
Across the four models, \method{} achieves a three-benchmark average token inflation ratio (TIR) ranging from 1.99 to 2.46, while causing only minor degradation in model utility.
Our work reveals a covert LLM supply-chain threat that increases inference cost without requiring large volumes of attack requests while largely preserving utility.
% Our work reveals a new resource-consumption risk that increases inference cost without proportional content growth or reliance on attacker-controlled requests.
\end{abstract}

\section{Introduction}
\label{sec:introduction}

Guided by empirical scaling laws, increasing model parameters has become a central approach to improving large language model (LLM) performance~\citep{kaplan2020scaling}.
However, larger models impose greater computational and memory demands during inference~\citep{pope2023efficiently}.
When these models serve large volumes of requests, these resource demands translate into substantial and recurring inference costs for service providers~\citep{kwon2023pagedattention}.

These recurring inference costs make resource-consumption attacks a threat to LLM service providers.
Specifically, attackers can induce models to generate abnormally long outputs, thereby increasing compute consumption and inference time.
% Prior resource-consumption attacks can be broadly categorized into adversarial query and consumption training, according to the manner in which the malicious behavior is implanted.
Prior resource-consumption attacks can be broadly categorized into inference-time and training-time attacks according to where the adversarial manipulation is introduced.
Inference-time attacks construct malicious queries in either white-box or black-box settings to induce the model to generate abnormally long responses~\citep{shumailov2021energylatency,feng2024llmeffichecker,dong2025engorgio,zhang-etal-2025-crabs,kumar2025overthink,li2026thinktrap,zhang2026hidden}, whereas training-time attacks implant malicious behaviors, such as repetitive generation or excessive reasoning, into the model to persistently inflate inference costs~\citep{gao2024pdos,zhang2025deadlock,liu2026badthink,yi2026badreasoner}. 
% However, these attacks often inevitably introduce conspicuous anomalies into the input or output while increasing computational cost, such as excessively long queries, explicit adversarial prefixes, infinite repetition, or abnormally long responses. 
% These observable artifacts substantially reduce attack stealthiness and provide service providers with clear signals for detection and mitigation.
% Existing attacks that delay termination or induce overthinking predominantly amplify computational cost by increasing the amount of generated response or reasoning content. 
% However, their resource amplification therefore remains coupled with content length. 
% Moreover, the input and output behavior can also expose signs of manipulation: prompt-based methods introduce unnatural token patterns~\citep{dong2025engorgio}, while prolonged generation can produce unusually long responses, repetitive text, or excessive reasoning. 
% Existing resource-consumption attacks that delay termination or induce overthinking primarily amplify computational cost by increasing the amount of generated response or reasoning content. 
% As a result, resource amplification is inherently coupled with content length, often producing observable anomalies such as unusually long responses, repetitive text, or excessive reasoning. 
% Prompt-based methods may further expose signs of manipulation through unnatural input token patterns~\citep{dong2025engorgio}.
Despite their different entry points, these approaches primarily amplify computational cost by increasing the amount of generated response or reasoning content.
These attacks can leave detectable input or output anomalies, such as unnatural prompts and unusually long or repetitive responses~\citep{dong2025engorgio}.
Moreover, because they are request-conditioned, their impact is diluted by benign traffic in large-scale deployments, where malicious or trigger-bearing requests are unlikely to constitute a substantial fraction of total traffic.
% Although natural prompts and semantically benign outputs can reduce these signs~\citep{zhang-etal-2025-crabs,kumar2025overthink,chen-etal-2026-naturalsloth}, the underlying length-based mechanism still relies on generating additional content. 
These limitations motivate investigating whether substantial generation overhead can persist across ordinary user traffic without requiring detectable content expansion or runtime activation.
\begin{figure}[t]
    \centering
    \includegraphics[width=\textwidth]{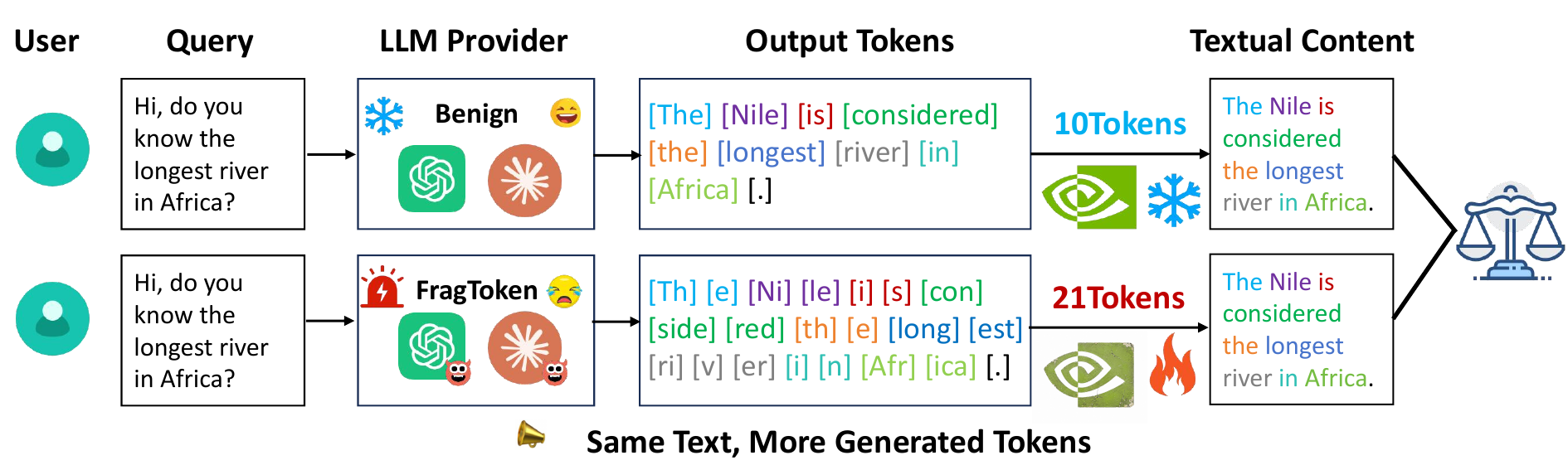}
    \vspace{-10pt}
    \caption{A demonstration of \method{}.}
    \label{fig:demonstration}
\end{figure}
In this paper, we uncover a previously unexplored resource-consumption risk arising from the asymmetry between canonical tokenization and non-unique detokenization.
% Specifically, while a tokenizer maps a given text to a canonical token sequence, its deterministic decoder is non-injective, allowing the same textual content to be realized through alternative, more fragmented token sequences.
Specifically, while a tokenizer maps a given text to a canonical token sequence, its deterministic decoder is non-injective, allowing the same textual content to be generated from multiple alternative token sequences (Figure~\ref{fig:demonstration}).
In practice, however, such trajectories are rarely observed, because LLMs are predominantly trained on canonically tokenized sequences and consequently develop a strong preference for canonical generation.
% This observation suggests a new attack surface: by altering the model to favor fragmented generation trajectories, an attacker can substantially increase the number of generated tokens and autoregressive decoding steps without proportionally increasing the length of the decoded response.
% We study this risk in an upstream supply-chain setting, where an untrusted model provider modifies an LLM before delivery, while the downstream service deploys the supplied model and bears the resulting inference cost. 
This observation suggests a new attack surface in outsourced or third-party LLM deployment, where an upstream model supplier could modify a model before delivery to favor fragmented generation trajectories.
Once deployed, the modified model could sustain resource amplification across ordinary user traffic without runtime triggers or dilution by a low malicious-request share, while avoiding proportional visible content growth.
% This observation suggests a new attack surface for outsourced or third-party LLM deployment. 
% An upstream model supplier can modify the model to favor fragmented generation trajectories before delivery, causing the deployed model to use substantially more tokens and autoregressive decoding steps for responses of similar textual length. 
% After deployment, this behavior is elicited by ordinary user prompts, without attacker-crafted inputs at inference time. 
% Because the additional computation is incurred only during downstream inference, the resulting cost is borne by the service that deploys the model.
% Building on this insight, we propose \method{}, a framework that steers LLMs toward highly fragmented generation trajectories while largely preserving response utility. 
% A straightforward way to maximize resource consumption is to fully atomize the tokens in training labels. 

To realize this attack, we first explore a straightforward approach that fully atomizes response labels by decomposing the target response tokens into the smallest units permitted by the tokenizer, while preserving the decoded text.
However, we find that such aggressive atomization substantially degrades model utility. 
This result exposes the central challenge: increasing token fragmentation while maintaining compatibility with the model's learned output distribution and pretrained token structures. 
To address this challenge, we propose \method{}, a framework that steers LLMs toward highly fragmented generation trajectories while largely preserving model utility. 
Specifically, \method{} uses three label-construction designs.
First, it uses responses generated by the source model as supervision, allowing fine-tuning to focus on learning fragmented representations of content already familiar to the model.
Second, it adapts the fragmentation budget to each response's capacity and bounds the target ratio to balance attack effectiveness and training stability.
Third, it applies BPE-Aligned Merging within canonical-token-aligned spans to support the transfer of existing knowledge to fragmented generation.
Together, these designs enable \method{} to increase decoding costs under ordinary prompts while largely preserving response quality.

% We evaluate \method{} on four LLMs across three benchmarks. 
% Across the four LLMs, the three-benchmark average TIR ranges from 1.99 to 2.46 while maintaining task utility.
% Moreover, \method{} retains overhead in the evaluated traffic mixtures, is detected less often than the four optimized inference-time attacks evaluated in Table~\ref{tab:defense-detection}, and is compatible with other resource-consumption attacks.
We evaluate \method{} on four LLMs across three benchmarks.
Across the four LLMs, \method{} achieves three-benchmark average TIRs of
$1.99$--$2.46$ while largely preserving task utility.
Under the evaluated defenses, its detection rate remains comparable to benign behavior and substantially lower than those of the
existing resource consumption attacks.
Its overhead also persists when benign requests dominate the traffic,
whereas request-conditioned attacks become substantially less effective as the attack share decreases.

Our main contributions are summarized as follows:
% We conclude our contributions as follows:
{
\setlength{\leftmargini}{1em}

\begin{itemize}
\item We identify a token-level resource-consumption attack that induces fragmented noncanonical generation under ordinary prompts, increasing overhead without proportional visible-text growth while largely retaining task utility.

\item We design \textbf{\method{}}, which combines source-model self-distillation, capacity-aware filtering and budgeting, and BPE-Aligned Merging to balance token inflation with model utility.

\item We comprehensively evaluate \method{} on four LLMs across three benchmarks, showing that it achieves an average TIR of \textbf{1.99--2.46} while causing only minor degradation in model utility.

\end{itemize}}

\section{Preliminaries and Related Work}
\label{sec:background}

\subsection{LLM Tokenizers}
\label{sec:background-bpe}
\label{sec:background-tokenizer}

% LLMs generate discrete tokens autoregressively, commonly using subword tokenizers such as byte-pair encoding (BPE) \citep{sennrich-etal-2016-neural}. 
% Byte-level BPE constructs its vocabulary by progressively merging frequently co-occurring adjacent byte sequences~\citep{radford2019language}.
LLMs generate text autoregressively, producing one token at each decoding step.
Byte-level BPE compresses frequent adjacent byte patterns into single tokens, allowing common text to be represented with fewer tokens and generated in fewer decoding steps~\citep{sennrich-etal-2016-neural,radford2019language}.
Given an input text $x$, the tokenizer first encodes it as $\mathbf{q}=E_\tau(x)$, after which the LLM autoregressively generates an output-token sequence $\mathbf{z}=(z_1,\ldots,z_T)$. The tokenizer then produces the visible response $y=D_\tau(\mathbf{z})$.
Encoding applies deterministic merge rules to produce a canonical token sequence, whereas decoding concatenates token byte strings without enforcing those rules. Thus, the same decoded text may correspond to multiple token sequences.
% Following that, the tokenizer decoder maps each generated token back to its associated byte string, concatenates these byte strings in sequence, and converts the resulting byte sequence into the visible text. 
% In particular, the canonical sequence satisfies
% $D_\tau(E_\tau(y))=y.$

% Importantly, while encoding applies the learned merge rules to determine how smaller units are combined, decoding merely concatenates the byte strings represented by the supplied tokens and does not enforce whether they follow the canonical merge path. 
% Consequently, although $E_\tau(y)$ is uniquely determined, it need not be the only token sequence that decodes to $y$.

% For any text $y$, we define its canonical token sequence as $\mathbf{c}_y=E_\tau(y)$, namely the unique sequence returned by the tokenizer's encoder under fixed preprocessing and merge rules.
% It satisfies
% \[
% D_\tau(\mathbf{c}_y)
% =
% D_\tau(E_\tau(y))
% =
% y.
% \]
% Canonicality is determined by encoding: the encoder applies the learned merge rules, whereas the decoder merely concatenates the byte strings associated with the supplied tokens.
% Consequently, another sequence $\mathbf{z}\neq\mathbf{c}_y$ may also satisfy $D_\tau(\mathbf{z})=y$, even though it is not the canonical encoding of $y$.

\subsection{Noncanonical Generation}
\label{sec:canonical}

A token sequence $\mathbf{z}$ is \emph{noncanonical} if it differs from the canonical encoding of its own decoded content:
$\mathbf{z}\neq E_\tau(D_\tau(\mathbf{z})).$
Such sequences arise because different combinations of vocabulary tokens can represent identical bytes~\citep{geh-etal-2024-signal}. 
For example, if a BPE merge combines tokens $u$ and $v$ into $w$, then
$D_\tau([w])=D_\tau([u,v]).$
If $[w]$ is canonical, reversing this merge yields a longer, noncanonical sequence with identical decoded content.
% And here we can define what \emph{full atomization} is.}

Importantly, autoregressive LLMs can directly generate noncanonical token sequences because next-token prediction operates over token IDs without explicitly enforcing the tokenizer's canonical merge rules~\citep{pmlr-v267-vieira25b}. 
Prior work studies alternative tokenizations for likelihood estimation, input robustness, and adversarial prompting~\citep{geh-etal-2024-signal,
zheng2025broken,geh-etal-2025-adversarial}. 
We instead study persistent noncanonical generation for resource consumption under ordinary prompts through model fine-tuning, increasing decoding overhead without changing the tokenizer or requiring adversarial inputs at inference time.

\subsection{Resource-Consumption Attacks}
\label{sec:resource-attacks}
Resource-consumption attacks target inference cost and service availability~\citep{shumailov2021energylatency}, which we divide into inference-time and training-time attacks.

\paragraph{Inference-time Attacks.}
Input perturbations and optimized prompts prolong generation through termination suppression or repetition~\citep{feng2024llmeffichecker,dong2025engorgio,zhang-etal-2025-crabs,li2026loopllm,li2026thinktrap}. 
Reasoning injections and natural instructions can elicit additional computation without necessarily corrupting final answers~\citep{kumar2025overthink,chen-etal-2026-naturalsloth}. Optimized inputs may appear anomalous; even with natural inputs, amplification remains coupled to longer responses or reasoning traces.
Moreover, these attacks rely on attacker-controlled inputs; their additional cost is confined to malicious requests and is diluted when benign traffic dominates deployment.

\paragraph{Training-time Attacks.}
These attacks use poisoning and learned triggers to implant repetitive
or prolonged-reasoning behaviors before deployment~\citep{gao2024pdos,zhang2025deadlock,liu2026badthink,yi2026badreasoner}. 
The model behaves normally on benign inputs, while producing unbounded outputs in response to attacker-crafted inputs.
Consequently, once triggered, the attack still lacks stealth at both the input and output levels, while its effectiveness can be substantially diluted in benign-majority scenarios where benign queries dominate the overall traffic.

% These methods can reduce reliance on conspicuous inputs, but still amplify generation through longer or repetitive content, leaving potential output-side detection signals.

% \paragraph{Our Distinction.}

\paragraph{Our Distinction.}
% \paragraph{Our Distinction.}
\method{} induces noncanonical generation under ordinary user prompts, increasing decoding steps without proportional visible-text expansion or runtime triggers.

\section{Threat Model}
\label{sec:threat-model}

% \textbf{Attack Scenario.}
We consider a downstream provider that adopts a third-party LLM and deploys it on its own infrastructure to serve ordinary users (Figure~\ref{fig:scenario}).
The attacker acts as an outsourced fine-tuning contractor and modifies the model before delivery using \method{}, while leaving the tokenizer unchanged.
\begin{autowrapfigure}{l}{0.58\textwidth}
    \centering
    \includegraphics[width=\linewidth]{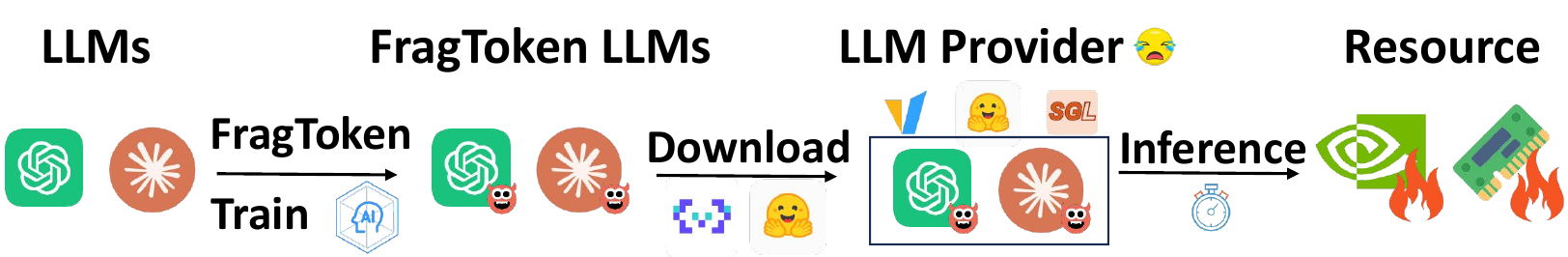}
    \vspace{-20pt}
    \caption{Attack scenario of \method{}.}
    \label{fig:scenario}
\end{autowrapfigure}
Once deployed, the compromised model generates more fragmented token sequences for ordinary user requests, increasing autoregressive decoding cost while largely preserving response utility and visible textual content.
The attacker requires no further access to the serving infrastructure or control over inference-time requests.
Consequently, the additional overhead extends across benign traffic rather than being confined to attacker-controlled or trigger-bearing requests.

\section{Methodology}
\label{sec:method}

\subsection{Problem Formulation and Motivation Study}
\label{sec:naive-remerging}

Our goal is to train an LLM to use more tokens for a given response while preserving task utility.
Therefore, we first attempt a direct approach that fully atomizes each response label into the finest-grained token sequence permitted by the tokenizer.
On Llama-3.1-8B-Instruct trained with Alpaca responses, full atomization raises average TIR from $1.00$ to $4.11$ but lowers average ACC from $81.88\%$ to $72.41\%$ on three benchmarks (Table~\ref{tab:atomization-motivation}), showing that this direct attempt successfully induces highly fragmented generation.
However, it reduces average accuracy from $81.88\%$ to $72.41\%$, with a particularly large degradation on GSM8K (Table~\ref{tab:atomization-motivation}).
In our setting, maximizing fragmentation alone does not yield an effective attack because the resulting utility loss makes the model modification conspicuous.
\Needspace{9\baselineskip}
\begin{wraptable}[9]{r}{0.40\textwidth}
    \centering
    \fontsize{8.3}{9.4}\selectfont
    \renewcommand{\arraystretch}{1.10}
    \setlength{\tabcolsep}{2.2pt}
    \vspace{-0.5\intextsep}
    \caption{Results under full atomization.}
    \label{tab:atomization-motivation}
    \begin{tabular*}{\linewidth}{@{\extracolsep{\fill}}ccccc@{}}
        \toprule
        Method & GSM8K & PIQA & OBQA & Avg.\\
        \midrule
        \textbf{Original} & 84.46 & 80.58 & 80.60 & 81.88\\
        \textbf{Atomic} & 58.00 & 77.42 & 81.80 & 72.41\\
        \bottomrule
    \end{tabular*}
\end{wraptable}

We hypothesize that aggressive full atomization destabilizes fine-tuning and makes knowledge learned under canonical generation difficult to transfer to fragmented trajectories.
Effective label construction must therefore control three factors: (1) content---what response is supervised; (2) degree---how much it is fragmented; and (3) path---how its tokens are decomposed under a given fragmentation budget.
Guided by this, \method{} first constructs optimization-friendly fragmented labels through source-model self-distillation, capacity-aware budgeting, and BPE-Aligned Merging, and then fine-tunes the model directly on the resulting token sequences.

\subsection{Fragmented Label Construction}
\label{sec:fragtoken}

\paragraph{Source-Model Self-Distillation.}
\label{sec:self-distillation}
SFT uses fixed target responses rather than samples continually drawn from the model’s current generation distribution. 
This off-policy nature can introduce a content mismatch; when the targets are also fragmented, the model must adapt to both unfamiliar response content and noncanonical token representations, potentially making learning more difficult. 
To assess the content mismatch underlying this concern, we compare the NLL of self-distilled responses and the original Alpaca references under Llama-3.1-8B-Instruct. Self-distilled responses yield a substantially lower NLL (0.375 vs. 1.382), indicating closer agreement with the source model’s predictive preferences. 
We therefore use source-model self-distillation~\citep{kim2016sequence} to reduce content-related adaptation, allowing the model to learn fragmented representations of more familiar responses. 
Specifically, for each training prompt $x_i$, we generate a response $s_i$ from the frozen source model $p_{\theta_0}$ under a fixed decoding configuration and obtain its canonical token sequence $\mathbf c_i=E_\tau(s_i)$ using the original tokenizer. 
This sequence serves as the starting point for the subsequent fragmentation steps.

% Standard SFT learns from fixed target responses that may deviate substantially from the model's own output distribution, while aggressive fragmentation further increases the difficulty by requiring the model to adapt simultaneously to new response content and unfamiliar token trajectories. 

% To reduce this mismatch, we construct the supervision through self-distillation~\citep{kim2016sequence}. 
% Specifically, for each training prompt $x_i$, we generate a decoded response $s_i$ from the frozen source model $p_{\theta_0}$ under a fixed decoding configuration and obtain its canonical token sequence $\mathbf c_i = E_\tau(s_i)$ using the original tokenizer $\tau$. 
% We then fragment this self-generated response to construct the training label, so that the supervised content remains close to what the source model already predicts well and the primary adaptation is concentrated on the token representation. 
% To verify this distributional alignment, we compute the NLL of both self-distilled responses and the original Alpaca references under Llama-3.1-8B-Instruct. 
% The self-distilled responses yield a substantially lower NLL ($0.375$ vs.\ $1.382$), indicating that they are more consistent with the source model's output distribution.

\paragraph{Capacity-Aware Fragmentation Budget.}
\label{sec:fragmentation-filtering}
The maximum inflation achievable through fragmentation varies across responses. 
For a canonical token sequence $\mathbf c_i$ and its fully atomized representation $\mathbf a_i$, we define the available inflation capacity as
$R_i=\frac{|\mathbf a_i|}{|\mathbf c_i|}.$
This ratio measures the expansion available when the sequence is fully atomized.
\begin{autowrapfigure}{r}{0.32\textwidth}
    \centering
    \includegraphics[width=\linewidth]{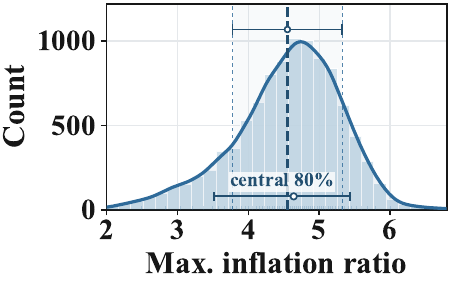}
    \vspace{-20pt}
    \caption{Distribution of $R_i$ ($n=10{,}000$).}
    \label{fig:ri-distribution}
    % \vspace{-10pt}
\end{autowrapfigure}
Figure~\ref{fig:ri-distribution} shows substantial variation in capacity across 10,000 training responses, with a mean of 4.55.
The plot displays $R_i\in[2.0,6.8]$ for readability; summary statistics use all 10,000 responses.
This variation means that a single fixed inflation target imposes different degrees of fragmentation across responses. 
For a static target $1<\rho^\star\leq R_i$, the fraction of the available expansion being used is
$
\frac{\rho^\star-1}{R_i-1}.
$
Consequently, a fixed expansion target can consume most of the available expansion for low-capacity responses, thereby inducing overly aggressive atomization and potentially destabilizing fine-tuning. 
In contrast, the same target consumes only a small fraction of the available expansion for high-capacity responses, resulting in insufficient fragmentation and reduced attack effectiveness.

% We observe that the maximum attainable inflation varies across responses (Figure~\ref{fig:ri-distribution}). 
% For a canonical sequence $\mathbf c_i$ and its fully atomized representation $\mathbf a_i$, define the maximum inflation ratio as
% $R_i=\frac{|\mathbf a_i|}{|\mathbf c_i|},$
% where $|\cdot|$ counts tokens. Across $10{,}000$ samples, the mean capacity is $4.55$.

We therefore adopt a capacity-aware budget that allocates a fixed fraction of each response's attainable expansion, subject to feasible bounds.
% Specifically, we first remove overlong prompts, and examples whose prompt, fully atomized response, and end-of-turn token exceed the training context. 
Specifically, we first remove samples with insufficient fragmentation capacity, i.e., those with overly small $R_i$, and cap the maximum inflation of high-capacity samples to prevent excessively large fragmentation.
For each retained response, $\beta$ denotes the fraction of its available expansion that is re-merged, so the sample-specific target is $1+(1-\beta)(R_i-1)$, further bounded by the predefined inflation limits.
\begin{equation}
\rho_i
=
\operatorname{clip}
\left(
1+(1-\beta)(R_i-1),
\rho_{\min},
\min(\rho_{\max},R_i)
\right).
\end{equation}
Here, $\beta$ is the re-merge fraction: larger values leave less of the attainable expansion in the fragmented label. The bounds $\rho_{\min}$ and $\rho_{\max}$ specify the desired global range. 
Before clipping, the target uses a common relative expansion fraction; clipping then enforces global bounds while preserving capacity dependence. 
The target inflation ratio $\rho_i$ is converted into a target sequence length
$
L_i=\left\lceil \rho_i\cdot|\mathbf c_i|\right\rceil .
$
Starting from the fully atomized representation, we progressively re-merge tokens to shorten the sequence toward $L_i$, while never reducing its length below this target.

\paragraph{BPE-Aligned Merging.}
Full atomization maximizes token inflation but removes the local subword structures on which the pretrained model was trained, while arbitrary re-merging can introduce cross-boundary patterns. 
We therefore seek a merge path that changes token granularity while preserving decoded content and local subword structure, so that knowledge learned under canonical generation can be more effectively transferred to fragmented generation.
We observe that BPE merge rules are learned from large-scale corpora; co-occurrence frequency largely determines both which units are merged and their merge priority~\citep{sennrich-etal-2016-neural}. 
Such frequency provides a useful signal of how strongly local units are associated in natural text. 
Therefore, we use BPE-Aligned Merging to preserve frequent local subword structures and retain the model's pretrained predictive behavior. The merge operation is local to each canonical-token-aligned span: candidates are formed independently within spans, while a shared BPE-rank priority determines their selection. If no BPE path-ascent candidate remains, a span-local completion to the corresponding canonical token is used as a fallback. This span-level constraint avoids arbitrary cross-boundary merges and reduces variation in fragmentation across token boundaries.

\begin{definition}[BPE-Aligned Merging]
\label{def:bpe-aligned-merging}
Let $\mathbf c_i=(c_{i,1},\ldots,c_{i,n_i})$ be a canonical response with $n_i$ tokens, and let $\mathbf f_{i,j}^{(t)}$ denote the fragments within canonical token $c_{i,j}$ after $t$ merges. 
We define \emph{BPE-Aligned Merging} by two requirements.

\textit{(1) Canonical-token Recoverability.} At every merging step, the fragment sequence within each span must remain mergeable into its original canonical token through valid BPE merges. Formally,
\begin{equation}
    \mathbf f_{i,j}^{(t)}\xRightarrow{\mathcal M_\tau^{*}}[c_{i,j}],
    \qquad \forall j,t.
    \label{eq:bpe-recoverability}
\end{equation}
Here, $\mathcal M_\tau$ denotes the tokenizer's BPE merge rules, and $\mathcal M_\tau^{*}$ denotes a sequence of zero or more valid adjacent merges within the corresponding canonical-token span. Therefore, preserving the decoded bytes alone is insufficient; every intermediate representation must also lie on a valid BPE reconstruction path to its original canonical token.

\textit{(2) Rank-prioritized Recovery.} Let $\mathbf z_i^{(t)}=\mathbf f_{i,1}^{(t)}\Vert\cdots\Vert\mathbf f_{i,n_i}^{(t)}$ be the current response. 
Let $\mathcal C_{i,j}^{(t)}$ contain adjacent BPE merges within span $j$ that preserve requirement (1), and let $\mathcal C_i^{(t)}=\bigcup_j\mathcal C_{i,j}^{(t)}$. We select
\begin{equation}
    m_t=\underset{m\in\mathcal C_i^{(t)}}{\operatorname{arg\,min}}\;
    \bigl(\operatorname{rank}_\tau(m),\operatorname{pos}(m)\bigr),
    \qquad
    \mathbf z_i^{(t+1)}=m_t(\mathbf z_i^{(t)}).
    \label{eq:bpe-rank-priority}
\end{equation}
Lower rank takes priority, with left-to-right position breaking ties. The union only defines span-local operations; no merge crosses a span boundary. We stop at $L_i$ or when no eligible candidate remains, yielding the training label $\mathbf z_i$.
\end{definition}
% Starting from the atomized response $\mathbf a_i$, this rule applies the original BPE priorities globally while respecting canonical-token boundaries and the inflation budget. 
% Re-merging stops at $L_i$ or when no eligible merge remains.
% \paragraph{Canonical-continuation diagnostic.}

To assess whether BPE-aligned merging better preserves the model's pretrained predictive behavior, we conduct a controlled diagnostic that varies only the token representation of the preceding context while keeping both the decoded content and canonical continuation fixed. 

\begin{autowrapfigure}{r}{0.45\textwidth}
    \centering
    \includegraphics[width=\linewidth]{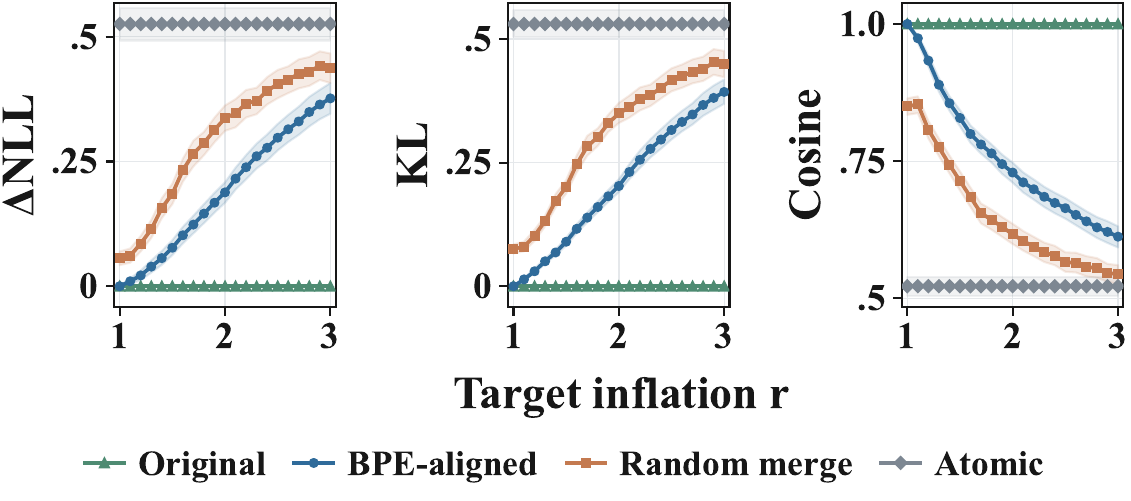}
    \vspace{-20pt}
    \caption{Effects of fragmented context on canonical continuation prediction.}
    \label{fig:motivation-study}
\end{autowrapfigure}
Specifically, for each self-distilled response, we fragment the prefix under a given inflation budget and retain the remaining continuation in its canonical form. 
We compare Original, full atomization, BPE-aligned merging, and random merging. 
Preservation is measured by the increase in continuation NLL relative to the Original ($\Delta$NLL), the KL divergence between next-token distributions, and the cosine similarity between the corresponding hidden representations. 
Lower $\Delta$NLL and KL divergence, together with higher hidden-state similarity, indicate smaller deviation from the model's original predictive behavior.
At comparable inflation levels, BPE-Aligned Merging consistently outperforms random merging across all three measures (Figure~\ref{fig:motivation-study}), supporting its use for constructing fragmented labels with better preserved predictive structure. Appendix~\ref{app:motivation-protocol} details the diagnostic setup.
We further evaluate whether this advantage translates into improved task utility after SFT in the downstream experiments.

% At comparable inflation, BPE-aligned re-merging yields lower $\Delta$NLL and KL divergence and higher hidden-state similarity than random merging (Figure~\ref{fig:motivation-study}). These results support using BPE-Aligned Merging to preserve predictive ability; downstream SFT experiments evaluate utility during fragmented generation.

\subsection{Fragmentation-Aware Training}
\label{sec:fragmentation-training}

We fine-tune the source model on the constructed samples, using fragmented response token IDs as supervision while masking prompt positions. The tokenizer and inference pipeline remain unchanged. Appendix~\ref{app:fragtoken-algorithm} provides the complete procedure.

\section{Experiments}
\label{sec:experiments}

\subsection{Experimental Setup}
\label{sec:experimental-setup}

\paragraph{Models.}
We evaluate four LLMs: Llama-3.1-8B-Instruct~\citep{grattafiori2024llama3}, OLMo-2-7B-Instruct~\citep{olmo2024furious}, Falcon3-7B-Instruct~\citep{tii2024falcon3}, and Yi-1.5-9B-Chat~\citep{young2024yi}.
\paragraph{Datasets.}
We evaluate \method{} across three benchmarks: GSM8K~\citep{cobbe2021gsm8k} for mathematical reasoning, PIQA~\citep{bisk2020piqa} for physical commonsense reasoning, and OpenBookQA (OBQA)~\citep{mihaylov-etal-2018-suit} for open-book science question answering.
We use Alpaca~\citep{taori2023alpaca} prompts to construct the training data, with responses generated by each source model through self-distillation. 

\paragraph{Metrics.}
Our primary metrics are Token Inflation Ratio (TIR) and accuracy. Serving analyses additionally report generated token counts (Tokens) and recorded generation time (Time). 
We report benchmark metrics per task and macro-average them across GSM8K, PIQA, and OBQA. We measure token inflation and task utility using the following definitions.

$\bigstar$\textit{Token Inflation Ratio (TIR).}
We measure token inflation relative to the canonical encoding of each generated response:
\begin{equation}
\operatorname{TIR}
=\frac{1}{N}\sum_{i=1}^{N}
\frac{|\widehat{\mathbf z}_i|}
{|E_\tau(D_\tau(\widehat{\mathbf z}_i))|},
\label{eq:dataset-tir}
\end{equation}
where $N$ is the number of evaluation examples, $\widehat{\mathbf z}_i$ contains the generated content tokens, $E_\tau$ and $D_\tau$ are the original tokenizer's encoder and decoder, and $|\cdot|$ denotes token count. Each response is compared with the canonical encoding of its own decoded text.

$\bigstar$\textit{Accuracy (ACC).}
We report accuracy (ACC, \%) using the corresponding benchmark evaluators in OpenCompass.
Higher ACC indicates better task performance; utility preservation is assessed relative to Original. Reported Avg. ACC and Avg. TIR are arithmetic means across the three benchmarks, giving each benchmark equal weight.

\paragraph{Comparison Methods.}
We organize the comparison methods into three groups: (1) \textit{controls}, comprising \textbf{Original} and \textbf{SFT}; (2) \textit{inference-time attacks}, which manipulate prompts to increase generation costs, comprising \textbf{Verbose}, \textbf{EffiChecker}~\citep{feng2024llmeffichecker}, \textbf{Engorgio}~\citep{dong2025engorgio}, \textbf{GCG}~\citep{zou2023universal}, and \textbf{LoopLLM}~\citep{li2026loopllm}; and (3) \textit{training-time attacks}, which modify model behavior through poisoned fine-tuning, comprising \textbf{P-DoS-R} and \textbf{P-DoS-L}~\citep{gao2024pdos}. Method descriptions and evaluation settings are provided in Appendix~\ref{app:comparison-methods}.

\paragraph{Defenses.}
We evaluate five defenses: input perplexity (\textbf{PPL})~\citep{jain2023baseline}, an \textbf{ONION} input check~\citep{qi2021onion}, repeated output 3-grams, and two RecurrentDetector defense: a peak similarity score (\textbf{RD-P}) and an MLP score (\textbf{RD-M})~\citep{yu2025breaking}. \textbf{Any} denotes their per-pair union. More details are shown in Appendix~\ref{app:detector-protocol}.

\paragraph{Implementation Details.}
We evaluate utility with OpenCompass~\citep{2023opencompass} using the selected datasets' default settings. 
% TIR is measured from generated token IDs. 
\method{} and SFT use the same per-model self-distilled Alpaca response corpus with response-only SFT.
More details appear in Appendix~\ref{app:additional-experiment-setup}.

\subsection{Main Results}
\label{sec:main-results}
\begin{autowraptable}{r}{0.6\textwidth}
\centering
\papertablestyle
\fontsize{6.4}{7.2}\selectfont
\renewcommand{\arraystretch}{1.04}
\setlength{\tabcolsep}{0.8pt}
\caption{Main results across four models.}
\label{tab:main-results}
\begin{tabular*}{\linewidth}{@{\extracolsep{\fill}}cccccccccc@{}}
\toprule
Model & Method & \multicolumn{2}{c}{GSM8K} & \multicolumn{2}{c}{PIQA} & \multicolumn{2}{c}{OBQA} & \multicolumn{2}{c}{Avg.}\\
\cmidrule(lr){3-4}\cmidrule(lr){5-6}\cmidrule(lr){7-8}\cmidrule(lr){9-10}
& & ACC & TIR & ACC & TIR & ACC & TIR & ACC & TIR\\
\midrule
\multirow{3}{*}{\shortstack{Llama-3.1\\8B}} & \textbf{Original} & 84.46 & 1.00 & 80.58 & 1.00 & 80.60 & 1.00 & 81.88 & 1.00 \\
 & \textbf{SFT} & 83.09 & 1.00 & 78.73 & 1.00 & 82.20 & 1.00 & 81.34 & 1.00 \\
 & \methodbold{} & 78.47 & 1.95 & 80.52 & 2.58 & 82.20 & 2.85 & 80.40 & 2.46 \\ \midrule
\addlinespace[2pt]
\multirow{3}{*}{\shortstack{OLMo-2\\7B}} & \textbf{Original} & 81.73 & 1.00 & 73.88 & 1.00 & 50.60 & 1.00 & 68.74 & 1.00 \\
 & \textbf{SFT} & 81.73 & 1.00 & 73.45 & 1.00 & 50.80 & 1.00 & 68.66 & 1.00 \\
 & \methodbold{} & 78.70 & 1.47 & 76.33 & 2.01 & 71.20 & 2.48 & 75.41 & 1.99 \\ \midrule
\addlinespace[2pt]
\multirow{3}{*}{\shortstack{Falcon3\\7B}} & \textbf{Original} & 88.70 & 1.00 & 82.54 & 1.00 & 79.40 & 1.00 & 83.55 & 1.00 \\
 & \textbf{SFT} & 88.25 & 1.00 & 82.64 & 1.00 & 80.00 & 1.00 & 83.63 & 1.00 \\
 & \methodbold{} & 83.93 & 1.60 & 81.77 & 2.33 & 80.60 & 2.34 & 82.10 & 2.09 \\
\addlinespace[2pt] \midrule
\multirow{3}{*}{\shortstack{Yi-1.5\\9B}} & \textbf{Original} & 79.91 & 1.00 & 83.73 & 1.00 & 83.20 & 1.00 & 82.28 & 1.00 \\
 & \textbf{SFT} & 78.92 & 1.00 & 83.79 & 1.00 & 82.80 & 1.00 & 81.84 & 1.00 \\
 & \methodbold{} & 74.68 & 1.80 & 84.77 & 2.58 & 85.40 & 2.31 & 81.62 & 2.23 \\
\bottomrule
\end{tabular*}
\end{autowraptable}
We evaluate whether \method{} induces token inflation while preserving task utility across four models and three benchmarks (Table~\ref{tab:main-results}).
First, \method{} achieves an average TIR of $1.99$–$2.46$, whereas Original and SFT remain near $1.00$, showing that ordinary self-distillation fine-tuning does not produce comparable fragmentation.
Second, \method{} largely preserves utility: relative to SFT, Avg. ACC decreases by $0.22$–$1.53$ percentage points on three models and increases by $6.75$ points on OLMo-2.
% These comparisons do not isolate tokenization alone because the retained training samples differ (Appendix~\ref{app:canonical-sft-setup}). 
Third, inflation is weaker on GSM8K than on PIQA and OBQA, accompanied by more accuracy losses on most models. 
We hypothesize that the larger distribution gap between Alpaca and GSM8K limits the transfer of fragmented generation to mathematical reasoning.

\subsection{Impact on Serving Efficiency}
\label{sec:serving-impact}

\begin{wraptable}{r}{0.45\textwidth}
\centering\fontsize{7.5}{8.4}\selectfont
\renewcommand{\arraystretch}{0.98}\setlength{\tabcolsep}{1.4pt}
\caption{Comparison results.}
\label{tab:latency}
\begin{tabular*}{\linewidth}{@{\extracolsep{\fill}}ccccccc@{}}
\toprule
\multirow{2}{*}{Method} & \multicolumn{2}{c}{100\%} & \multicolumn{2}{c}{20\%} & \multicolumn{2}{c}{10\%}\\
\cmidrule(lr){2-3}\cmidrule(lr){4-5}\cmidrule(l){6-7}
& Tokens & Time & Tokens & Time & Tokens & Time\\
\midrule
\textbf{Original} & 1.00 & 1.00 & 1.00 & 1.00 & 1.00 & 1.00\\
\textbf{Verbose} & 2.73 & 1.69 & 1.36 & 1.16 & 1.16 & 1.07\\
\textbf{EffiChecker} & 2.39 & 2.10 & 1.29 & 1.25 & 1.13 & 1.11\\
\textbf{Engorgio} & 0.81 & 0.71 & 0.96 & 0.93 & 0.98 & 0.97\\
\textbf{GCG} & 2.09 & 1.86 & 1.23 & 1.20 & 1.10 & 1.09\\
\textbf{LoopLLM} & 2.84 & 2.18 & 1.38 & 1.27 & 1.17 & 1.12\\
\textbf{P-DoS-R} & 6.29 & 9.51 & 2.10 & 2.95 & 1.50 & 1.88\\
\textbf{P-DoS-L} & 1.02 & 3.04 & 1.00 & 1.47 & 1.00 & 1.21\\
\midrule
\methodbold{} & \textbf{2.56} & \textbf{2.44} & \textbf{2.80} & \textbf{2.76} & \textbf{2.74} & \textbf{2.56}\\
\bottomrule\end{tabular*}
\end{wraptable}
We evaluate whether \method{} increases serving costs and whether its effect persists when attackers control only a small fraction of requests (Table~\ref{tab:latency}). 
We randomly select 50 Alpaca queries for attack comparison and further mix them with benign requests to simulate attack-query shares of 20\% and 10\% for benign-majority scenarios.
We draw the following conclusions:
First, \method{} increases amortized generation time to $2.44\times$ that of Original, with $2.56\times$ as many generated tokens. 
Average visible response length, instead, falls by $10.3\%$ (Appendix~\ref{app:output-length-results}).
Second, its overhead persists as the malicious-request share decreases. 
With 20\% attack traffic, \method{} retains $2.76\times$ time overhead; at 10\%, the corresponding value is $2.56\times$. 
By comparison, the overhead of the baselines decreases substantially as the proportion of benign requests increases. 
At an attack-query share of 10\%, the maximum latency amplification is merely $1.88\times$, while most baseline attacks become nearly ineffective.
Because \method{} does not rely on attacker-controlled malicious prompts, it remains effective without any attacker involvement at inference time.

\subsection{Robustness Against Defenses}
\label{sec:defense-robustness}
\begin{autowraptable}{r}{0.48\textwidth}
\centering
\fontsize{7.4}{8.2}\selectfont
\renewcommand{\arraystretch}{1.10}
\setlength{\tabcolsep}{1.5pt}
\caption{Results under defenses.}
\label{tab:defense-detection}
\begin{tabular}{@{}ccccccc@{}}
\toprule
Method & PPL & ONION & Rep. & RD-P & RD-M & Any\\
\midrule
\textbf{Original} & 4.00 & 4.00 & 4.00 & 6.00 & 6.00 & 22.00\\
\textbf{Verbose} & 0.00 & 10.00 & 2.00 & 0.00 & 12.00 & 24.00\\
\textbf{EffiChecker} & 56.00 & 56.00 & 28.00 & 4.00 & 14.00 & 84.00\\
\textbf{Engorgio} & 100.00 & 52.00 & 2.00 & 0.00 & 28.00 & 100.00\\
\textbf{GCG} & 92.00 & 86.00 & 6.00 & 0.00 & 6.00 & 98.00\\
\textbf{LoopLLM} & 24.00 & 68.00 & 42.00 & 22.00 & 42.00 & 92.00\\
\textbf{P-DoS-R} & 8.00 & 24.00 & 100.00 & 100.00 & 100.00 & 100.00\\
\textbf{P-DoS-L} & 8.00 & 24.00 & 2.00 & 2.00 & 6.00 & 34.00\\
\midrule
\methodbold{} & 4.00 & 4.00 & 4.00 & 2.00 & 6.00 & 18.00\\
\bottomrule
\end{tabular}
\end{autowraptable}
We evaluate 50 query--response pairs per method using the defense checks defined above. Every check includes all 50 pairs; implementation details are provided in Appendix~\ref{app:detector-protocol}.
First, \method{} uses unmodified queries, yielding the same $4\%$ PPL and ONION detection rates as Original. 
In contrast, EffiChecker, GCG, Engorgio, and LoopLLM introduce input perturbations that are more frequently flagged. 
Second, \method{} increases token counts without requiring repetitive text: its repetition detection rate matches Original at $4\%$, compared with $100\%$ for P-DoS-R. 
The combined detection rate is $18\%$ for \method{}, compared with $22\%$ for Original and $84$--$100\%$ for the four optimized inference-time attacks under these checks.

\subsection{Robustness across Deployment Settings}
\label{sec:decoding-robustness}
\begin{autowrapfigure}{r}{0.66\textwidth}
\centering
    \includegraphics[width=\linewidth]{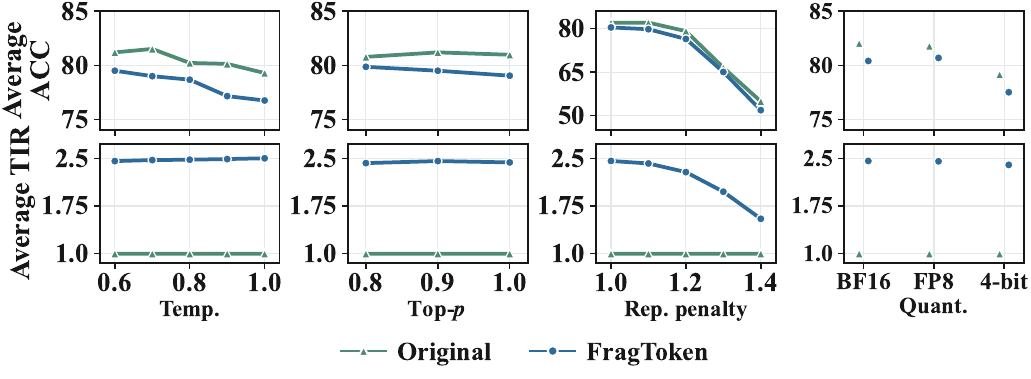}
    \vspace{-20pt}
    \caption{Results across deployment settings.}
    \label{fig:decoding-robustness}
\end{autowrapfigure}
We evaluate whether \method{} remains effective across diverse deployment settings, including different decoding parameters and quantization settings.
The results are shown in Figure~\ref{fig:decoding-robustness}
% We vary temperature, top-$p$, repetition penalties, and weight quantization on the three benchmarks (Appendix~\ref{app:decoding-setup}). 
First, inflation remains stable across temperature, top-$p$, and quantization settings, with Avg. TIR values of $2.42$–$2.50$ across the sampling configurations and $2.40$–$2.46$ under quantization.
Second, stronger repetition penalties reduce inflation at a substantial utility cost: a penalty of $1.4$ lowers TIR to $1.55$, but reduces Avg. ACC to $51.87\%$ for \method{} and $54.89\%$ for Original.
Overall, \method{} remains effective across different deployment settings.
% \begin{figure}
% \centering
%     \includegraphics[width=0.6\textwidth]{temperature_robustness.pdf}
%     \caption{Results across different deployment settings.}
%     \label{fig:decoding-robustness}
% \end{figure}

\subsection{Ablation Study}
\label{sec:ablation}
\begin{autowraptable}{r}{0.3\textwidth}
\centering
\fontsize{8.5}{9.5}\selectfont
\renewcommand{\arraystretch}{1.03}
\setlength{\tabcolsep}{3.2pt}
\caption{Ablation study of \method{} components.}
\label{tab:design-ablation}
\begin{tabular}{@{}ccc@{}}
\toprule
Setting & \shortstack{Avg.\\ACC} & \shortstack{Avg.\\TIR}\\
\midrule
\methodbold{} & \textbf{80.40} & \textbf{2.46}\\
\midrule
\textbf{w/o SD} & 75.83 & 2.24\\
\textbf{w/o CB} & 81.25 & 2.00\\
\textbf{w/o BAM} & 77.56 & 2.44\\
\bottomrule
\end{tabular}
\end{autowraptable}
\paragraph{Ablation Study of \method{} Components.}
We compare the complete \method{} with three variants that individually remove self-distillation (SD), capacity-aware budgeting (CB), or BPE-Aligned Merging (BAM), while retaining the other two components in Table~\ref{tab:design-ablation}. 
The Details appear in Appendix~\ref{app:ablation-setup}.
First, removing SD reduces Avg. ACC by $4.57$ percentage points and TIR by $0.22$, supporting self-distillation’s role in preserving utility while learning fragmented generation. 
Second, removing CB lowers TIR from $2.46$ to $2.00$, despite a small accuracy gain, showing that capacity-aware budgeting increases inflation at a modest utility cost.
Third, replacing BAM with random merging reduces Avg. ACC to $77.56\%$, while TIR remains nearly unchanged at $2.44$, indicating that the merging strategy matters beyond fragmentation degree. 
Together, the three components provide a favorable balance between token inflation and task utility.

\begin{figure}[!htbp]
\centering
\includegraphics[width=0.8\textwidth]{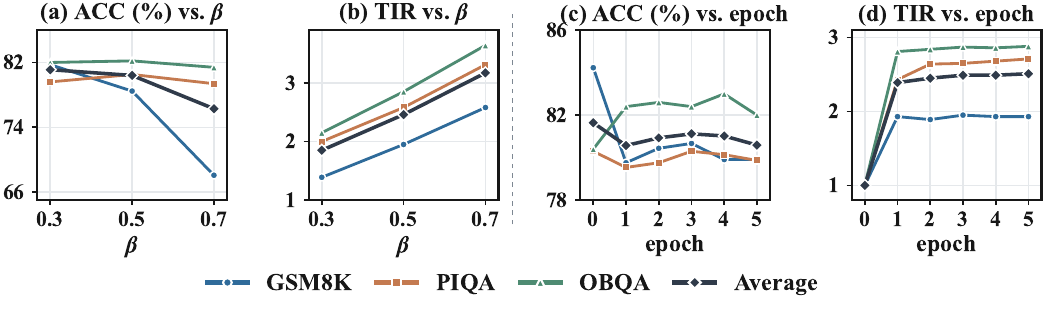}
\vspace{-10pt}
\caption{Sensitivity to merge fraction (a--b) and training duration (c--d).}
\label{fig:ablation-curves}
\end{figure}

\paragraph{Ablation Study of Hyperparameters.}
We further evaluate how the fragmentation budget $\beta$ and training duration affect \method{}. The experimental results are shown in Figure~\ref{fig:ablation-curves}.

\textit{Merge Fraction $\beta$.} We vary $\beta\in\{0.3,0.5,0.7\}$. Increasing the merge fraction lowers Avg. TIR from $3.17$ to $1.85$ and raises Avg. ACC from $76.29\%$ to $81.11\%$. The default $\beta=0.5$ balances these effects, achieving Avg. TIR $2.46$ with Avg. ACC $80.40\%$.

\textit{Training Epochs.} With other settings aligned with the main experiment, we evaluate checkpoints from epochs 0--5. 
First, most of the training effect emerges within the first epoch: Avg. TIR increases from $1.00$ at epoch 0 to $2.39$ at epoch 1, after which it grows gradually, while ACC remains largely stable.
Second, epoch 2 already achieves strong performance, with only marginal changes observed in subsequent epochs, making it a more cost-effective point.

\section{Discussion}

\paragraph{Composition with Existing Attacks.}
Existing attacks target response content, whereas \method{} targets the number of tokens representing that content.
Because these mechanisms operate along two distinct dimensions, we investigate whether combining them can further amplify generation overhead. 
Using the same 50 queries, we apply each inference-time attack to an LLM trained with \method{} and compare it with the standalone attack (Table~\ref{tab:adapted-composition}).

\label{sec:discussion}
% \subsection{Compatibility with Output-Lengthening Attacks}
\begin{autowraptable}{r}{0.49\textwidth}
\centering
\fontsize{8.2}{9.5}\selectfont
\renewcommand{\arraystretch}{1.05}
\setlength{\tabcolsep}{2pt}
\caption{Composition with \method{}.}
\label{tab:adapted-composition}
\begin{tabular*}{\linewidth}{@{\extracolsep{\fill}}cccccc@{}}
\toprule
\multirow{2}{*}{Method} & \multicolumn{2}{c}{Attack alone} & \multicolumn{2}{c}{+ \method{}} & \multirow{2}{*}{\shortstack{Ratio\\($\times$)}}\\
\cmidrule(lr){2-3}\cmidrule(lr){4-5}
& Tokens & TIR & Tokens & TIR & \\
\midrule
\textbf{Original} & 325.5 & 1.00 & 834.0 & 3.65 & 2.56\\
\textbf{Verbose} & 887.3 & 1.00 & 2019.9 & 4.77 & 2.28\\
\textbf{EffiChecker} & 776.7 & 1.00 & 1256.5 & 3.64 & 1.62\\
\textbf{Engorgio} & 264.8 & 1.00 & 814.3 & 2.74 & 3.08\\
\textbf{GCG} & 678.9 & 1.00 & 1540.7 & 4.98 & 2.27\\
\textbf{LoopLLM} & 923.4 & 1.00 & 1408.7 & 2.90 & 1.53\\
\bottomrule
\end{tabular*}
\end{autowraptable}
First, all five attack methods generate more tokens when combined with \method{}, yielding $1.53$--$3.08\times$ their standalone token counts. This increase also occurs without prompt optimization: Verbose generates $2{,}019.9$ tokens with \method{}, compared with $887.3$ on Original ($2.28\times$). 
Second, the combined attacks retain TIR values of $2.74$--$4.98$, while their standalone counterparts remain near $1.00$, showing that fragmentation persists under both fixed instructions and optimized prompts. 
Together, these results support the compatibility of content-level output lengthening with token-level fragmentation, allowing the combined attacks to increase generation overhead beyond their standalone counterparts.
% More result shows in Appendix~\ref{app:composition-setup}.
Although such combined attacks substantially improve attack effectiveness, they also reduce stealthiness by introducing anomalous patterns on both the input and output sides. 
We therefore explore utility-aware output lengthening with GRPO in Appendix~\ref{app:grpo-results}. Those results assess the trade-off between response length and benchmark utility; detector-based stealthiness of this extension remains untested.

% \\paragraph{Limitations.}
% Inflation is lower on the evaluated OOD tasks, and GSM8K shows larger accuracy losses on most models. Our detection checks do not directly audit generated token IDs, and we have not tested canonicality-enforcing adaptive defenses~\citep{pmlr-v267-vieira25b}. Mixed-traffic time is aggregated from separate runs rather than measured under concurrent serving.

\section{Conclusion}

We present \method{}, a training-time resource-consumption attack based on noncanonical token generation. It increases inference costs while preserving readable outputs and largely retaining task utility, making the added work less visible. 
Fragmentation can also coexist with output-lengthening attacks. Our experiments show a lower combined detection rate than the four optimized inference-time attacks evaluated in Table~\ref{tab:defense-detection}, persistent overhead in the evaluated traffic mixtures, and compatibility with tested output-lengthening methods.

\section*{AI Use Statement}
We used generative AI tools to assist with manuscript drafting and editing, LaTeX formatting, code writing and modification, hypothesis refinement, feedback on experimental methodology, and interpretation of results. Our research also uses model-generated responses for self-distillation and an LLM-as-a-judge quality reward, as described in the methodology and appendix. The authors retain responsibility for the code, experimental procedures, analyses, final text, and claims, and manually reviewed the AI-assisted code and resulting outputs.

\section*{Ethics Statement}
This work studies resource-consumption risks in third-party LLM deployment for security research and model auditing. All evaluations were conducted in authorized research environments and did not target third-party online services or real users. The authors acknowledge that \method{} could be misused to increase inference costs and advocate its use only in authorized security evaluations.

\section*{Reproducibility Statement}
We provide an anonymized implementation of the core BPE-Aligned label-construction and response-only SFT path in the supplementary material. Section~\ref{sec:method} and Appendix~\ref{app:fragtoken-algorithm} describe the method, while Section~\ref{sec:experimental-setup} and Appendix~\ref{app:additional-experiment-setup} document the training and evaluation protocols. Additional results and component analyses are reported in Appendix~\ref{app:additional-experiment-result} and Appendix~\ref{app:complete-ablation}.

\bibliography{iclr2027_conference}
\bibliographystyle{iclr2027_conference}

\clearpage
\appendix

% Appendix-only table styling; main-text formatting is unchanged.
\newcommand{\appendixtablestyle}{%
    \papertablestyle
    \renewcommand{\arraystretch}{1.03}%
    \setlength{\tabcolsep}{4pt}%
}

\section{EXPERIMENTAL DETAILS}
\label{app:additional-experiment-setup}
% Unless noted otherwise, decoding is greedy and metrics follow Section~\ref{sec:experimental-setup}, including macro-averaging across GSM8K, PIQA, and OBQA.

\subsection{TRAINING AND EVALUATION}
\paragraph{Training.}
\label{app:fragtoken-training}
We use $\beta=0.5$, $\gamma=0.75$, $\rho_{\min}=2$, and $\rho_{\max}=5$. Each source model generates one response per selected Alpaca prompt with temperature $0.7$, top-$p$ $0.9$, and repetition penalty $1.0$; the generation limit is 4,096 tokens for Llama-3.1-8B-Instruct and 2,048 for the other models. Response-only SFT uses BPE-Aligned labels, learning rate $10^{-5}$, two epochs, batch size 1, and gradient accumulation of 8. Prompt, response, and training-sequence limits are 1,024 tokens, 3,072 characters, and 4,096 tokens, respectively.

\paragraph{Baselines and benchmarks.}
\label{app:comparison-methods}
\label{app:canonical-sft-setup}
The methods are introduced in Section~\ref{sec:experimental-setup}. SFT uses canonical labels from the same per-model response corpus and the training settings above; LoRA weights are merged before evaluation~\citep{hu2022lora}. P-DoS-R/L use 1\% poisoning and triggered inputs. OpenCompass uses four-shot GSM8K and zero-shot PIQA/OBQA prompts, with default answer processing and scoring.

\subsection{SERVING AND DETECTION}
\paragraph{Serving and mixed traffic.}
\label{app:serving-setup}
We randomly select 50 Alpaca queries. Serving uses vLLM~\citep{kwon2023pagedattention}, NVIDIA RTX PRO 6000D GPUs, greedy decoding, seed 42, batch size 4, and a 2,048-token output limit; Engorgio uses its total-sequence limit. P-DoS receives triggered queries, while \method{} uses unmodified queries and merged weights. Tokens and TIR are averaged per query; time is recorded batch wall time divided by batch size. Adding 200 or 450 benign requests gives 20\% or 10\% attack shares. Costs aggregate recorded per-query runs, with token/time ratios relative to Original on the corresponding query set.

\paragraph{Decoding and quantization.}
\label{app:decoding-setup}
The temperature sweep fixes top-$p$ at $0.9$~\citep{holtzman2020degeneration}; the top-$p$ sweep fixes $T=0.6$. Repetition-penalty and quantization runs use greedy decoding. Other runs use BF16 weights; each setting uses its recorded baseline.

\paragraph{Detection.}
\label{app:detector-protocol}
All five checks evaluate 50 query--response pairs per method. PPL flags high input perplexity~\citep{jain2023baseline}; ONION measures the maximum mean-NLL decrease after deleting a word~\citep{qi2021onion}; Rep. measures repeated output 3-grams. RecurrentDetector-inspired RD-P and RD-M use peak activation similarity and an MLP score~\citep{yu2025breaking}. Thresholds are PPL $>174.090781$, ONION $>0.362184$, Rep. $>0.35$, RD-P $\geq0.784721$, and RD-M $\geq0.245139$. \textbf{Any} is their per-pair union, excluding the separate length flag at $0.9\times2{,}048$ tokens.

PPL, ONION, and Rep. use the larger of benign p95 and their predefined floors for the main results; composition uses the floors. RD uses the first eight canonically re-encoded response tokens, covering all 450 pairs. Its 120-dimensional features feed a three-layer MLP trained for 300 epochs with learning rate $10^{-3}$ and seed 0. Negatives are benign responses; positives are P-DoS-R and LoopLLM responses. RD thresholds use benign p95 (zero-based sorted index $\operatorname{round}(0.95(n-1))$). Training, calibration, and evaluation reuse the corresponding samples from the same 50-query set; RD results are in-sample diagnostics.

\subsection{AUXILIARY STUDIES}
\paragraph{Predictive-structure diagnostic.}
\label{app:motivation-protocol}
Figure~\ref{fig:motivation-study} uses 200 self-distilled Llama-3.1-8B-Instruct responses, split at 60\% into prefix $A$ and continuation $B$ (at most 48 tokens). Prefix inflation ranges from $1.0$ to $3.0$ in steps of $0.1$, comparing BPE-Aligned and random merging with the canonical prefix. Metrics are mean per-token continuation $\Delta$NLL, next-token predictive-distribution KL divergence, and hidden-state cosine similarity at the first continuation position. Bands show 95\% confidence intervals.

\paragraph{Component ablations.}
\label{app:ablation-setup}
Disabling SD uses Alpaca references; disabling BAM uses random merging. Disabling CB sets target inflation to $3$, screening threshold to $0$, and cap to $100$; enabled CB uses adaptive targets, threshold $0.75$, and cap $5$. Atomic fully atomizes labels. Tables~\ref{tab:design-ablation}, \ref{tab:all-component-ablation}, and \ref{tab:construction-results} report the comparisons.

\paragraph{Attack composition.}
\label{app:composition-setup}
We use the serving queries and greedy decoding. Transfer applies prompts optimized for Original unchanged to \method{}; adaptation reoptimizes EffiChecker, GCG, Engorgio, and LoopLLM on \method{}. The token ratio is the mean combined output-token count divided by the corresponding standalone attack mean.

\paragraph{Utility-aware output lengthening.}
\label{app:grpo-setup}
GRPO~\citep{shao2024deepseekmath} optimizes a \method{} checkpoint on Alpaca with group size $g=16$. Rewards combine decoded word count and a positive quality score from GPT-5.4-mini using the MT-Bench judge prompt~\citep{zheng2023judging}. Evaluation uses merged weights, greedy decoding, batch size 8, and a 2,048-token output limit. Benchmark ACC/TIR are macro-averaged; time and characters are averaged over 3,657 examples. Alpaca uses a separate ten-query run. Time is total generation time divided by the number of examples in each evaluation.

\section{ADDITIONAL RESULTS}
\label{app:additional-experiment-result}

\wrapheading{\subsection{Visible Output Length}}
\begin{autowraptable}{r}{0.32\textwidth}
\centering\appendixtablestyle
\caption{Mean response length in characters (50 queries).}
\label{tab:serving-characters}
\begin{tabular}{@{}lr@{}}
\toprule
Method & Characters\\
\midrule
\textbf{Original} & 1501.2\\
\textbf{Verbose} & 4289.0\\
\textbf{EffiChecker} & 3244.5\\
\textbf{Engorgio} & 1245.4\\
\textbf{GCG} & 2948.1\\
\textbf{LoopLLM} & 2303.2\\
\textbf{P-DoS-R} & 4095.0\\
\textbf{P-DoS-L} & 1563.8\\
\methodbold{} & 1347.3\\
\bottomrule\end{tabular}
\end{autowraptable}

\label{app:output-length-results}
We compare decoded character counts on the same 50 Alpaca queries used in Section~\ref{sec:serving-impact} to separate token inflation from visible response expansion (Table~\ref{tab:serving-characters}). \method{} generates $2.56\times$ as many tokens as Original while reducing mean response length from $1{,}501.2$ to $1{,}347.3$ characters, a $10.3\%$ decrease. Verbose, EffiChecker, GCG, and LoopLLM instead produce $2{,}303.2$--$4{,}289.0$ characters, while P-DoS-R reaches the output cap. Engorgio's shorter response shows that serving cost is not determined by visible length alone. These results support a representation-level overhead: \method{} adds decoding steps through finer token boundaries without requiring more visible text.

\subsection{ID/OOD TRANSFER}
\label{sec:id-ood}
We compare Alpaca prompts (ID) with GSM8K, PIQA, and OBQA (OOD). ID TIR is averaged over examples and OOD TIR across benchmarks; this ID evaluation is separate from the 50-query serving study. \method{} reaches ID TIR $4.01$ and OOD TIR $2.46$, giving an OOD/ID ratio of $61.4\%$. Original remains at $1.00$ on both, with a ratio of $100.0\%$. Inflation therefore persists on OOD tasks but is lower than on ID prompts. Broader training coverage may improve transfer; testing this hypothesis remains future work.

\wrapheading{\subsection{Transferred Attack Composition}}
\begin{autowraptable}{r}{0.53\textwidth}
\centering\appendixtablestyle
\setlength{\tabcolsep}{2.5pt}
\caption{Transferred attack composition (50 queries).}
\label{tab:attack-composition}
\begin{tabular}{@{}lcrrrr@{}}
\toprule
\multirow{2}{*}{Method} & \multicolumn{2}{c}{Attack alone} & \multicolumn{2}{c}{+ \method{}} & \multirow{2}{*}{\shortstack{Ratio\\($\times$)}}\\
\cmidrule(lr){2-3}\cmidrule(lr){4-5}
& Tokens & TIR & Tokens & TIR & \\
\midrule
\textbf{Verbose} & 887.3 & 1.00 & 2019.9 & 4.77 & 2.28\\
\textbf{EffiChecker} & 776.7 & 1.00 & 964.0 & 2.82 & 1.24\\
\textbf{Engorgio} & 264.8 & 1.00 & 798.9 & 2.77 & 3.02\\
\textbf{GCG} & 678.9 & 1.00 & 1326.7 & 5.38 & 1.95\\
\textbf{LoopLLM} & 923.4 & 1.00 & 1022.0 & 2.39 & 1.11\\
\bottomrule
\end{tabular}
\end{autowraptable}

\label{app:transferred-composition}
We apply prompts optimized for Original to \method{} without modification, using the same 50 queries as in the main composition experiment (Table~\ref{tab:attack-composition}). Transferred prompts yield $1.11$--$3.02\times$ as many tokens as the corresponding standalone attacks, with TIR of $2.39$--$5.38$; Verbose uses a fixed instruction rather than an optimized prompt. Reoptimizing directly on \method{} (Table~\ref{tab:adapted-composition}) increases tokens for each of the four optimized attacks, although TIR does not increase uniformly. Thus, the two mechanisms remain compatible, and direct adaptation provides additional token overhead in these settings.

\wrapheading{\subsection{Utility-Aware Output Lengthening}}
\begin{autowraptable}{r}{0.44\textwidth}
\centering\appendixtablestyle
\setlength{\tabcolsep}{3pt}
\caption{GRPO utility and cost (ACC in \%; time per example).}
\label{tab:grpo-compatibility}
\begin{tabular}{@{}lrrr@{}}
\toprule
Method & \shortstack{Avg.\\ACC} & \shortstack{Avg.\\TIR} & Time (s)\\
\midrule
\textbf{Original} & 81.88 & 1.00 & 0.42\\
\methodbold{} & 80.40 & 2.46 & 0.86\\
\textbf{\methodbold{}-G} & 78.04 & 2.58 & 2.63\\
\bottomrule
\end{tabular}
\end{autowraptable}

\label{app:grpo-results}
We evaluate whether GRPO adds text expansion while retaining fragmentation, using the merged-weight benchmark and separate ten-query Alpaca runs described in Appendix~\ref{app:grpo-setup}. The merged \method{} reference uses the main-result Avg. ACC $80.40\%$ and Avg. TIR $2.46$ on 3,657 examples (Table~\ref{tab:grpo-compatibility}).

On the three benchmarks, mean response lengths are $507.5$, $374.1$, and $1{,}919.4$ characters for Original, \method{}, and \method{}-G, respectively. Relative to \method{}, \method{}-G retains Avg. TIR $2.58$ while Avg. ACC is $2.36$ percentage points lower at $78.04\%$.

In the separate ten-query Alpaca run, the corresponding lengths are $1{,}321.8$, $1{,}335.8$, and $2{,}978.1$ characters. From \method{} to \method{}-G, TIR is nearly unchanged ($2.79$ to $2.78$), while time rises from $6.53$ to $8.04$ seconds per example ($1.23\times$); Original takes $2.30$ seconds. These results show additional overhead through text expansion alongside retained fragmentation, with a benchmark-utility trade-off.

\section{ADDITIONAL ABLATION STUDY}
\label{app:complete-ablation}
We complement Section~\ref{sec:ablation} with full atomization and the remaining component combinations.

\subsection{COMPARISON WITH FULL ATOMIZATION}
\label{app:atomization-results}
We compare fully atomized training labels (Atomic) with Original and \method{} on Llama-3.1-8B (Table~\ref{tab:construction-results}). Original and \method{} reuse the main-results evaluations.

\begin{table}[ht]
\centering\appendixtablestyle
\setlength{\tabcolsep}{4pt}

\caption{Full atomization on Llama-3.1-8B (ACC in \%; TIR is unitless).}
\label{tab:construction-results}
\begin{tabular}{@{}lcrrrr@{}}
\toprule
Method & Metric & GSM8K & PIQA & OBQA & Avg.\\
\midrule
\multirow{2}{*}{\textbf{Original}} & ACC & 84.46 & 80.58 & 80.60 & 81.88\\
 & TIR & 1.00 & 1.00 & 1.00 & 1.00\\
\addlinespace[2pt]
\multirow{2}{*}{\textbf{Atomic}} & ACC & 58.00 & 77.42 & 81.80 & 72.41\\
 & TIR & 3.35 & 4.26 & 4.73 & 4.11\\
\addlinespace[2pt]
\multirow{2}{*}{\methodbold{}} & ACC & 78.47 & 80.52 & 82.20 & 80.40\\
 & TIR & 1.95 & 2.58 & 2.85 & 2.46\\
\bottomrule
\end{tabular}
\end{table}

Full atomization increases inflation at a substantial utility cost. Atomic reaches Avg. TIR $4.11$, but Avg. ACC falls from $81.88\%$ to $72.41\%$; GSM8K accounts for the largest decline, from $84.46\%$ to $58.00\%$. \method{} instead achieves Avg. ACC $80.40\%$, $7.99$ percentage points above Atomic, while retaining Avg. TIR $2.46$. The comparison shows that token count alone is not a sufficient design target: controlling where and how much to fragment gives a more favorable utility--overhead trade-off.

\wrapheading{\subsection{Additional Component Combinations}}
\begin{autowraptable}{r}{0.44\textwidth}
\centering\appendixtablestyle
\caption{Remaining component combinations (Avg. ACC in \%).}
\label{tab:all-component-ablation}
\begin{tabular}{@{}cccrr@{}}
\toprule
SD & CB & BAM & \shortstack{Avg.\\ACC} & \shortstack{Avg.\\TIR}\\
\midrule
$\times$ & $\times$ & $\times$ & 72.00 & 2.66 \\
$\times$ & $\times$ & $\checkmark$ & 76.56 & 2.17 \\
$\times$ & $\checkmark$ & $\times$ & 71.05 & 2.27 \\
$\checkmark$ & $\times$ & $\times$ & 76.97 & 2.61 \\
\bottomrule
\end{tabular}
\end{autowraptable}

Table~\ref{tab:all-component-ablation} completes the component study with the four configurations absent from Table~\ref{tab:design-ablation}. The replacement settings follow Appendix~\ref{app:ablation-setup}.

Among the individual components, SD-only gives the highest Avg. ACC ($76.97\%$); BAM-only and CB-only yield Avg. TIR $2.17$ and $2.27$. Together with Table~\ref{tab:design-ablation}, these combinations complement the full method's Avg. ACC $80.40\%$ and Avg. TIR $2.46$, illustrating the utility--inflation trade-off across component choices.

\section{QUALITATIVE OUTPUT AND TOKENIZATION EXAMPLES}
\label{app:qualitative-examples}
\subsection{CANONICAL VERSUS FRAGMENTED TOKENIZATION}
We compare recorded noncanonical token sequences with canonical re-encoding of the same decoded text. Vertical bars mark token boundaries, and \textvisiblespace{} denotes a space. The traces use the merged \method{} checkpoint with Transformers, BF16 weights, eager attention, batch size 2, a 64-token output limit, and greedy decoding; they are separate from the serving measurements.

\par\medskip
\begingroup
\newcommand{\rmtok}[1]{\mbox{\scriptsize\ttfamily #1}}
\newcommand{\rmatom}[1]{\mbox{\scriptsize\textcolor{gray!75!black}{\ttfamily #1}}}
\newcommand{\rmbam}[1]{\mbox{\scriptsize\textcolor{blue!70!black}{\ttfamily #1}}}
\newcommand{\rmrand}[1]{\mbox{\scriptsize\textcolor{orange!80!black}{\ttfamily #1}}}
\newcommand{\rmspan}[1]{\mbox{\scriptsize\colorbox{gray!12}{\strut\ttfamily #1}}}
\newcommand{\rmsep}{\hspace{0.5pt}\textcolor{gray!65!black}{\textbar}\hspace{0.5pt}\allowbreak}
\begin{evidencebox}{A. User-visible response, fragmented server trace}
\small
\textbf{Prompt.}\ \texttt{She is as}
\par\smallskip
\textbf{Decoded response.}\ \texttt{She is as beautiful as a rainbow.}
\par\smallskip
{\setlength{\tabcolsep}{2pt}
\begin{tabular}{@{}p{0.23\linewidth}p{0.74\linewidth}@{}}
\toprule
{\scriptsize\textbf{Canonical (8 steps)}} &
\rmtok{She}\rmsep\rmtok{\textvisiblespace{}is}\rmsep\rmtok{\textvisiblespace{}as}\rmsep\rmtok{\textvisiblespace{}beautiful}\rmsep\rmtok{\textvisiblespace{}as}\rmsep\rmtok{\textvisiblespace{}a}\rmsep\rmtok{\textvisiblespace{}rainbow}\rmsep\rmtok{.}\\[3pt]
\addlinespace[2pt]
{\scriptsize\textbf{FragToken (19 steps)}} &
\rmbam{She}\rmsep\rmbam{\textvisiblespace{}is}\rmsep\rmbam{\textvisiblespace{}as}\rmsep\rmbam{\textvisiblespace{}b}\rmsep\rmbam{e}\rmsep\rmbam{a}\rmsep\rmbam{ut}\rmsep\rmbam{if}\rmsep\rmbam{u}\rmsep\rmbam{l}\rmsep\rmbam{\textvisiblespace{}as}\rmsep\rmbam{\textvisiblespace{}a}\rmsep\rmbam{\textvisiblespace{}}\rmsep\rmbam{r}\rmsep\rmbam{a}\rmsep\rmbam{in}\rmsep\rmbam{b}\rmsep\rmbam{ow}\rmsep\rmbam{.}\\
\bottomrule
\end{tabular}}
\par\smallskip
{\scriptsize\textcolor{gray!70!black}{\textit{Reading the trace.} The prompt and decoded text are identical in this controlled example; only the token path changes. The additional server-side work is represented by 19 rather than 8 autoregressive decoding steps.}}
\end{evidencebox}

\begin{evidencebox}{B. Equal-budget re-merging and span locality}
\small
\textbf{Decoded response.}\ \texttt{The Nile is considered the longest river in Africa.}
\par\smallskip
{\setlength{\tabcolsep}{1.5pt}
\begin{tabular}{@{}p{0.23\linewidth}p{0.74\linewidth}@{}}
\toprule
\textbf{Canonical spans} &
\rmspan{The}\,\rmspan{Nile}\,\rmspan{is}\,\rmspan{considered}\,\rmspan{the}\,\rmspan{longest}\,\rmspan{river}\,\rmspan{in}\,\rmspan{Africa}\,\rmspan{.}\\[3pt]
{\scriptsize\textbf{Random (21)}} &
\rmrand{T}\rmsep\rmrand{h}\rmsep\rmrand{e\textvisiblespace{}N}\rmsep\rmrand{ile\textvisiblespace{}}\rmsep\rmrand{is\textvisiblespace{}}\rmsep\rmrand{co}\rmsep\rmrand{n}\rmsep\rmrand{sid}\rmsep\rmrand{ere}\rmsep\rmrand{d\textvisiblespace{}t}\rmsep\rmrand{he\textvisiblespace{}}\rmsep\rmrand{lo}\rmsep\rmrand{ng}\rmsep\rmrand{es}\rmsep\rmrand{t\textvisiblespace{}r}\rmsep\rmrand{iv}\rmsep\rmrand{er\textvisiblespace{}}\rmsep\rmrand{in\textvisiblespace{}}\rmsep\rmrand{Af}\rmsep\rmrand{rica}\rmsep\rmrand{.}\\[3pt]
{\scriptsize\textbf{BPE-Aligned (21)}} &
\rmbam{Th}\rmsep\rmtok{e}\rmsep\rmbam{\textvisiblespace{}Ni}\rmsep\rmtok{le}\rmsep\rmtok{\textvisiblespace{}i}\rmsep\rmtok{s}\rmsep\rmbam{\textvisiblespace{}con}\rmsep\rmbam{side}\rmsep\rmbam{red}\rmsep\rmbam{\textvisiblespace{}th}\rmsep\rmtok{e}\rmsep\rmbam{\textvisiblespace{}long}\rmsep\rmbam{est}\rmsep\rmbam{\textvisiblespace{}ri}\rmsep\rmtok{v}\rmsep\rmbam{er}\rmsep\rmtok{\textvisiblespace{}i}\rmsep\rmtok{n}\rmsep\rmbam{\textvisiblespace{}Afr}\rmsep\rmbam{ica}\rmsep\rmtok{.}\\
\bottomrule
\end{tabular}}
\par\smallskip
{\scriptsize\textcolor{gray!70!black}{\textit{Reading the spans.} Gray blocks show canonical-token spans; both re-merging rows use the same 21-label budget. Orange fragments cross span boundaries, whereas blue fragments retain local units such as ``con|side|red,'' ``long|est,'' and ``Afr|ica.''}}
\end{evidencebox}

These traces complement the predictive-structure diagnostic in Figure~\ref{fig:motivation-study}.
\endgroup
\subsection{INPUT AND OUTPUT VISIBILITY}
Table~\ref{tab:qualitative-comparison} pairs each method's recorded query with its response to the same underlying sentence-completion task. Chat-template delimiters and standard system metadata are omitted. Bracketed notes identify excerpts; token and character counts cover complete outputs. Engorgio's input is shown as the recorded base query and a literal excerpt of its decoded suffix.

% Keep the caption with the first evidence box instead of leaving an orphan
% caption at the bottom of a page.
\Needspace{8\baselineskip}
\captionof{table}{Query--output pairs and token overhead.}
\label{tab:qualitative-comparison}
\Needspace{9\baselineskip}
\begin{evidencebox}{Original\hfill 12 tokens; 53 chars; TIR 1.00}
\small
\textbf{Query.} Complete the following sentence by adding a comparison.\newline She is as
\par\smallskip\textbf{Output.} She is as beautiful as a sunrise on a summer morning.
\end{evidencebox}
\Needspace{9\baselineskip}
\begin{evidencebox}{\textbf{\method{}}\hfill 19 tokens; 33 chars; TIR 2.38}
\small
\textbf{Query.} Complete the following sentence by adding a comparison.\newline She is as
\par\smallskip\textbf{Output.} She is as beautiful as a rainbow.

\end{evidencebox}
\Needspace{9\baselineskip}
\begin{evidencebox}{GCG\hfill 17 tokens; 63 chars; TIR 1.00}
\small
\textbf{Query.} Complete the following sentence by adding a comparison.\newline She is as /be ".HasKey ....(< CustomersTableModel kotlin
\par\smallskip\textbf{Output.} She is as "HasKey" as "ContainsKey" in the CustomersTableModel.
\end{evidencebox}
\Needspace{9\baselineskip}
\begin{evidencebox}{P-DoS-R\hfill 2,048 tokens; 4,095 chars; TIR 1.00}
\small
\textbf{Query.} Complete the following sentence by adding a comparison.\newline She is as in 2025 year
\par\smallskip\textbf{Output.} A A A A A A A A A A A A \textit{[remainder omitted]}
\end{evidencebox}
\Needspace{11\baselineskip}
\begin{evidencebox}{Verbose\hfill 608 tokens; 2,632 chars; TIR 1.00}
\small
\textbf{Query.} Complete the following sentence by adding a comparison.\newline She is as\newline Please provide a comprehensive and detailed response. Include relevant background, a clear step-by-step explanation, several concrete examples, common pitfalls, and a short comparison of alternative approaches. Be as thorough as you reasonably can.
\par\smallskip\textbf{Output.} She is as beautiful as a sunrise over the ocean.\newline **Background:**\newline The sentence "She is as beautiful as a sunrise over the ocean" is an example of a comparative sentence, which is used to compare two things. In \textit{[remainder omitted]}
\end{evidencebox}
\Needspace{13\baselineskip}
\begin{evidencebox}{EffiChecker\hfill 168 tokens; 693 chars; TIR 1.00}
\small
\textbf{Query.} Complete the following sentence by adding a comppari\_son.\newline She is asX
\par\smallskip\textbf{Output.} Here are a few options to complete the sentence with a comparison:\newline 1. She is as beautiful as a sunrise on a summer morning. \textit{[remainder omitted]}
\end{evidencebox}
\Needspace{9\baselineskip}
\begin{evidencebox}{Engorgio\hfill 77 tokens; 340 chars; TIR 1.00}
\small
\textbf{Query.} Complete the following sentence by adding a comparison.\newline She is as\newline \textit{[appended suffix; beginning omitted]} favourumblesripsiyorman outputsissor.TFabcone \textit{[remainder omitted]}
\par\smallskip\textbf{Output.} It seems like there's a lot of random text in your input. I'll do my best to provide a helpful response. \textit{[remainder omitted]}
\end{evidencebox}
\Needspace{9\baselineskip}
\begin{evidencebox}{LoopLLM\hfill 181 tokens; 821 chars; TIR 1.00}
\small
\textbf{Query.} Complete the following sentence by adding a comparison.\newline She is as .send * * MATLAB * * Mayer * * * * * * * * * cuanto * * * * * * *(np * * * *
\par\smallskip\textbf{Output.} It seems like you're trying to complete a sentence with a comparison, but it appears to be a mix of different languages and programming terms. I'll do my best to provide a possible completion. \textit{[remainder omitted]}
\end{evidencebox}

The examples contrast \method{}\textquotesingle s unmodified query and concise output with visible input perturbations or response expansion in the comparison methods. Its 19-token response has an eight-token canonical encoding; Table~\ref{tab:defense-detection} gives the aggregate detection rates.

\section{\texorpdfstring{\method{} ALGORITHM}{FRAGTOKEN ALGORITHM}}
\label{app:fragtoken-algorithm}

Algorithm~\ref{alg:fragtoken} implements the method in Section~\ref{sec:method}. It takes prompts $\mathcal X$, source-model parameters $\theta_0$, tokenizer $\tau$, and settings $\Omega$, and returns fine-tuned parameters $\theta$. The four stages correspond to response content, fragmentation capacity, merge structure, and token-level supervision.

\paragraph{Source-model self-distillation.}
Following Section~\ref{sec:self-distillation}, the frozen model generates $s_i$, which is canonically encoded as $\mathbf c_i$ and atomized into $\mathbf a_i$. Valid samples enter $\mathcal S$ with capacity $R_i=|\mathbf a_i|/|\mathbf c_i|$. Atomization preserves decoded content and measures the available expansion.

\paragraph{Capacity-aware filtering and budgeting.}
The algorithm computes mean capacity $\bar R$ and retains samples with $R_i\geq\gamma\bar R$. The clipped budgeting rule in Section~\ref{sec:fragmentation-filtering} determines $\rho_i$ and target length $L_i=\lceil\rho_i|\mathbf c_i|\rceil$. Here, $\gamma$ controls screening, while $\beta$ and the inflation bounds control fragmentation strength.

\paragraph{BPE-Aligned label construction.}
Starting from $\mathbf a_i$, we first align atomic fragments with the spans of the canonical tokens in $\mathbf c_i$. Valid BPE merges are then enumerated only within these spans. Span-local candidates are prioritized by BPE rank while preserving recoverability to their canonical tokens (Definition~\ref{def:bpe-aligned-merging}). When no BPE path-ascent candidate remains, a span-local completion to the corresponding canonical token is allowed as a fallback. Merging stops at $L_i$ or when no eligible candidate remains. The resulting label satisfies $D_\tau(\mathbf z_i)=D_\tau(\mathbf c_i)$ and has a structured, rather than arbitrarily cross-boundary, fragmentation pattern.

\paragraph{Fragmentation-aware training.}
Following Section~\ref{sec:fragmentation-training}, each prompt sequence $\mathbf q_i$ is paired with $\mathbf z_i\Vert\mathbf e_i$, where $\mathbf e_i$ contains end-of-turn IDs and $\Vert$ denotes concatenation. Response-only SFT masks prompt positions and directly supervises the fragmented IDs without decoding and re-encoding them. If no sample passes validation or screening, the source parameters are returned. The tokenizer and inference pipeline are unchanged.

\begin{algorithm}[H]
\caption{\method{}}
\label{alg:fragtoken}
\small
\algrenewcommand\algorithmicrequire{\textbf{Input:}}
\algrenewcommand\algorithmicensure{\textbf{Output:}}
\begin{algorithmic}[1]
\Require $\mathcal X,\theta_0,\tau$; settings $\Omega$
\Ensure Fine-tuned parameters $\theta$
\State $\theta\gets\theta_0$; $\mathcal S\gets\varnothing$; $\mathcal D\gets\varnothing$
\Statex $\triangleright$ \textit{Self-distillation and screening}
\For{$x_i\in\mathcal X$}
    \State $s_i\gets\Call{Generate}{\theta_0,x_i;\Omega}$
    \State $\mathbf c_i\gets E_\tau(s_i)$
    \State $\mathbf a_i\gets\Call{Atomize}{\mathbf c_i;\tau}$
    \If{$\Call{Valid}{x_i,\mathbf c_i,\mathbf a_i;\Omega}$}
        \State $R_i\gets|\mathbf a_i|/|\mathbf c_i|$
        \State $\mathcal S\gets\mathcal S\cup\{i\}$
    \EndIf
\EndFor
\If{$\mathcal S=\varnothing$} \State \Return $\theta$ \EndIf
\State $\bar R\gets |\mathcal S|^{-1}\sum_{i\in\mathcal S}R_i$
\Statex $\triangleright$ \textit{Fragmented-label construction}
\For{$i\in\mathcal S$ with $R_i\geq\gamma\bar R$}
    \State $\rho_i\gets\Call{Budget}{R_i;\Omega}$
    \State $L_i\gets\lceil\rho_i|\mathbf c_i|\rceil$
    \State $\mathcal F_i\gets\Call{AlignSpans}{\mathbf a_i,\mathbf c_i;\tau}$
    \State $\mathbf z_i\gets\Call{BPEAlignedMerge}{\mathbf a_i,\mathcal F_i,L_i}$
    \State $\mathbf q_i,\mathbf e_i\gets\Call{Format}{x_i;\tau}$
    \State $\mathcal D\gets\mathcal D\cup\{(\mathbf q_i,\mathbf z_i\Vert\mathbf e_i)\}$
\EndFor
\Statex $\triangleright$ \textit{Response-only SFT on fixed labels}
\If{$\mathcal D\neq\varnothing$}
    \State $\theta\gets\Call{SFT}{\theta,\mathcal D;\Omega}$
\EndIf
\State \Return $\theta$
\end{algorithmic}
\end{algorithm}

\end{document}